\documentclass[%
reprint,
superscriptaddress,
showpacs,
amsmath,amssymb,
aps,
prl,
]{revtex4-1}

\usepackage{graphicx}
\usepackage{dcolumn}
\usepackage{bm}
\usepackage{natbib}
\usepackage{multirow}

\usepackage{soul}
\setstcolor{red}
\usepackage{soul}
\soulregister\cite7 
\soulregister\citep7 
\soulregister\citet7 
\soulregister\ref7 
\soulregister\pageref7 

\usepackage{color}

\begin{document}

\preprint{APS/123-QED}

\title{
Search for Light Weakly-Interacting-Massive-Particle Dark Matter by Annual Modulation Analysis with 
a Point-Contact Germanium Detector at the China Jinping 
Underground Laboratory
}



\author{L.~T.~Yang}
\affiliation{Key Laboratory of Particle and Radiation Imaging 
(Ministry of Education) and Department of Engineering Physics, 
Tsinghua University, Beijing 100084}
\author{H.~B.~Li}
\affiliation{Institute of Physics, Academia Sinica, Taipei 11529~}

\author{Q. Yue}\altaffiliation 
[Corresponding author. ]{yueq@mail.tsinghua.edu.cn}
\affiliation{Key Laboratory of Particle and Radiation Imaging 
(Ministry of Education) and Department of Engineering Physics, 
Tsinghua University, Beijing 100084}
\author{H.~Ma}\altaffiliation 
[Corresponding author. ]{mahao@mail.tsinghua.edu.cn}
\affiliation{Key Laboratory of Particle and Radiation Imaging 
(Ministry of Education) and Department of Engineering Physics, 
Tsinghua University, Beijing 100084}
\author{K.~J.~Kang}
\affiliation{Key Laboratory of Particle and Radiation Imaging 
(Ministry of Education) and Department of Engineering Physics, 
Tsinghua University, Beijing 100084}
\author{Y.~J.~Li}
\affiliation{Key Laboratory of Particle and Radiation Imaging 
(Ministry of Education) and Department of Engineering Physics, 
Tsinghua University, Beijing 100084}

\author{H.~T.~Wong}
\affiliation{Institute of Physics, Academia Sinica, Taipei 11529~}

\author{M. Agartioglu}
\affiliation{Institute of Physics, Academia Sinica, Taipei 11529~}
\affiliation{Department of Physics, Dokuz Eyl\"{u}l University, 
\.{I}zmir 35160~}
\author{H.~P.~An}
\affiliation{Department of Physics, Tsinghua University, Beijing 100084}
\author{J.~P.~Chang}
\affiliation{NUCTECH Company, Beijing 100084}
\author{J.~H.~Chen}
\affiliation{Institute of Physics, Academia Sinica, Taipei 11529~}
\author{Y.~H.~Chen}
\affiliation{YaLong River Hydropower Development Company, Chengdu 610051}
\author{J.~P.~Cheng}
\affiliation{Key Laboratory of Particle and Radiation Imaging 
(Ministry of Education) and Department of Engineering Physics, 
Tsinghua University, Beijing 100084}
\affiliation{College of Nuclear Science and Technology, 
Beijing Normal University, Beijing 100875}
\author{Z.~Deng}
\affiliation{Key Laboratory of Particle and Radiation Imaging 
(Ministry of Education) and Department of Engineering Physics, 
Tsinghua University, Beijing 100084}
\author{Q.~Du}
\affiliation{College of Physical Science and Technology, 
Sichuan University, Chengdu 610064}
\author{H.~Gong}
\affiliation{Key Laboratory of Particle and Radiation Imaging 
(Ministry of Education) and Department of Engineering Physics, 
Tsinghua University, Beijing 100084}
\author{Q.~J.~Guo}
\affiliation{School of Physics, Peking University, Beijing 100871}
\author{L. He}
\affiliation{NUCTECH Company, Beijing 100084}
\author{J.~W.~Hu}
\affiliation{Key Laboratory of Particle and Radiation Imaging 
(Ministry of Education) and Department of Engineering Physics, 
Tsinghua University, Beijing 100084}
\author{Q.~D.~Hu}
\affiliation{Key Laboratory of Particle and Radiation Imaging 
(Ministry of Education) and Department of Engineering Physics, 
Tsinghua University, Beijing 100084}
\author{H.~X.~Huang}
\affiliation{Department of Nuclear Physics, 
China Institute of Atomic Energy, Beijing 102413}
\author{L.~P.~Jia}
\affiliation{Key Laboratory of Particle and Radiation Imaging 
(Ministry of Education) and Department of Engineering Physics, 
Tsinghua University, Beijing 100084}
\author{H. Jiang}
\affiliation{Key Laboratory of Particle and Radiation Imaging 
(Ministry of Education) and Department of Engineering Physics, 
Tsinghua University, Beijing 100084}
\author{H. Li}
\affiliation{NUCTECH Company, Beijing 100084}
\author{J.~M.~Li}
\affiliation{Key Laboratory of Particle and Radiation Imaging 
(Ministry of Education) and Department of Engineering Physics, 
Tsinghua University, Beijing 100084}
\author{J.~Li}
\affiliation{Key Laboratory of Particle and Radiation Imaging 
(Ministry of Education) and Department of Engineering Physics, 
Tsinghua University, Beijing 100084}
\author{X.~Li}
\affiliation{Department of Nuclear Physics, 
China Institute of Atomic Energy, Beijing 102413}
\author{X.~Q.~Li}
\affiliation{School of Physics, Nankai University, Tianjin 300071}
\author{Y.~L.~Li}
\affiliation{Key Laboratory of Particle and Radiation Imaging 
(Ministry of Education) and Department of Engineering Physics, 
Tsinghua University, Beijing 100084}

\author {B. Liao}
\affiliation{College of Nuclear Science and Technology, 
Beijing Normal University, Beijing 100875}

\author{F.~K.~Lin}
\affiliation{Institute of Physics, Academia Sinica, Taipei 11529~}
\author{S.~T.~Lin}
\affiliation{College of Physical Science and Technology, 
Sichuan University, Chengdu 610064}
\author{S.~K.~Liu}
\affiliation{College of Physical Science and Technology, 
Sichuan University, Chengdu 610064}

\author {Y.~D.~Liu}
\affiliation{College of Nuclear Science and Technology, 
Beijing Normal University, Beijing 100875}
\author {Y.~Y.~Liu}
\affiliation{College of Nuclear Science and Technology, 
Beijing Normal University, Beijing 100875}
\author{Z.~Z.~Liu}
\affiliation{Key Laboratory of Particle and Radiation Imaging 
(Ministry of Education) and Department of Engineering Physics, 
Tsinghua University, Beijing 100084}

\author{J.~L.~Ma}
\affiliation{Key Laboratory of Particle and Radiation Imaging 
(Ministry of Education) and Department of Engineering Physics, 
Tsinghua University, Beijing 100084}
\affiliation{Department of Physics, Tsinghua University, Beijing 100084}
\author{Y.~C.~Mao}
\affiliation{School of Physics, Peking University, Beijing 100871}
\author{H.~Pan}
\affiliation{NUCTECH Company, Beijing 100084}
\author{J.~Ren}
\affiliation{Department of Nuclear Physics, 
China Institute of Atomic Energy, Beijing 102413}
\author{X.~C.~Ruan}
\affiliation{Department of Nuclear Physics, 
China Institute of Atomic Energy, Beijing 102413}

\author{V.~Sharma}
\affiliation{Institute of Physics, Academia Sinica, Taipei 11529~}
\affiliation{Department of Physics, Banaras Hindu University, Varanasi 221005~}

\author{Z.~She}
\affiliation{Key Laboratory of Particle and Radiation Imaging 
(Ministry of Education) and Department of Engineering Physics, 
Tsinghua University, Beijing 100084}
\author{M.~B.~Shen}
\affiliation{YaLong River Hydropower Development Company, Chengdu 610051}
\author{L.~Singh}
\affiliation{Institute of Physics, Academia Sinica, Taipei 11529~}
\affiliation{Department of Physics, Banaras Hindu University, Varanasi 221005~}
\author{M.~K.~Singh}
\affiliation{Institute of Physics, Academia Sinica, Taipei 11529~}
\affiliation{Department of Physics, Banaras Hindu University, Varanasi 221005~}

\author {T.~X.~Sun}
\affiliation{College of Nuclear Science and Technology, 
Beijing Normal University, Beijing 100875}
\author{C.~J.~Tang}
\affiliation{College of Physical Science and Technology, 
Sichuan University, Chengdu 610064}
\author{W.~Y.~Tang}
\affiliation{Key Laboratory of Particle and Radiation Imaging 
(Ministry of Education) and Department of Engineering Physics, 
Tsinghua University, Beijing 100084}
\author{Y.~Tian}
\affiliation{Key Laboratory of Particle and Radiation Imaging 
(Ministry of Education) and Department of Engineering Physics, 
Tsinghua University, Beijing 100084}

\author {G.~F.~Wang}
\affiliation{College of Nuclear Science and Technology, 
Beijing Normal University, Beijing 100875}
\author{J.~M.~Wang}
\affiliation{YaLong River Hydropower Development Company, Chengdu 610051}
\author{L.~Wang}
\affiliation{Department of Physics, Beijing Normal University, Beijing 100875}
\author{Q.~Wang}
\affiliation{Key Laboratory of Particle and Radiation Imaging 
(Ministry of Education) and Department of Engineering Physics, 
Tsinghua University, Beijing 100084}
\affiliation{Department of Physics, Tsinghua University, Beijing 100084}
\author{Y.~Wang}
\affiliation{Key Laboratory of Particle and Radiation Imaging 
(Ministry of Education) and Department of Engineering Physics, 
Tsinghua University, Beijing 100084}
\affiliation{Department of Physics, Tsinghua University, Beijing 100084}
\author{Y.~X.~Wang}
\affiliation{School of Physics, Peking University, Beijing 100871}
\author{S.~Y.~Wu}
\affiliation{YaLong River Hydropower Development Company, Chengdu 610051}
\author{Y.~C.~Wu}
\affiliation{Key Laboratory of Particle and Radiation Imaging 
(Ministry of Education) and Department of Engineering Physics, 
Tsinghua University, Beijing 100084}
\author{H.~Y.~Xing}
\affiliation{College of Physical Science and Technology, 
Sichuan University, Chengdu 610064}
\author{Y.~Xu}
\affiliation{School of Physics, Nankai University, Tianjin 300071}
\author{T.~Xue}
\affiliation{Key Laboratory of Particle and Radiation Imaging 
(Ministry of Education) and Department of Engineering Physics, 
Tsinghua University, Beijing 100084}
\author{N.~Yi}
\affiliation{Key Laboratory of Particle and Radiation Imaging 
(Ministry of Education) and Department of Engineering Physics, 
Tsinghua University, Beijing 100084}
\author{C.~X.~Yu}
\affiliation{School of Physics, Nankai University, Tianjin 300071}
\author{H.~J.~Yu}
\affiliation{NUCTECH Company, Beijing 100084}
\author{J.~F.~Yue}
\affiliation{YaLong River Hydropower Development Company, Chengdu 610051}
\author{X.~H.~Zeng}
\affiliation{YaLong River Hydropower Development Company, Chengdu 610051}
\author{M.~Zeng}
\affiliation{Key Laboratory of Particle and Radiation Imaging 
(Ministry of Education) and Department of Engineering Physics, 
Tsinghua University, Beijing 100084}
\author{Z.~Zeng}
\affiliation{Key Laboratory of Particle and Radiation Imaging 
(Ministry of Education) and Department of Engineering Physics, 
Tsinghua University, Beijing 100084}

\author {F.~S.~Zhang}
\affiliation{College of Nuclear Science and Technology, 
Beijing Normal University, Beijing 100875}

\author{Y.~H.~Zhang}
\affiliation{YaLong River Hydropower Development Company, Chengdu 610051}
\author{M.~G.~Zhao}
\affiliation{School of Physics, Nankai University, Tianjin 300071}
\author{J.~F.~Zhou}
\affiliation{YaLong River Hydropower Development Company, Chengdu 610051}
\author{Z.~Y.~Zhou}
\affiliation{Department of Nuclear Physics, 
China Institute of Atomic Energy, Beijing 102413}
\author{J.~J.~Zhu}
\affiliation{College of Physical Science and Technology, 
Sichuan University, Chengdu 610064}
\author{Z.~H.~Zhu}
\affiliation{YaLong River Hydropower Development Company, Chengdu 610051}

\collaboration{CDEX Collaboration}


\date{\today}

\begin{abstract}
We present results on light weakly interacting massive particle (WIMP) searches with annual modulation 
(AM) analysis on data from a 1-kg mass $p$-type point-contact germanium detector of the CDEX-1B experiment 
at the China Jinping Underground Laboratory. Datasets with a total live time of 3.2 yr within a 
4.2 yr span are analyzed with analysis threshold of 250 eVee. Limits on WIMP-nucleus (${\chi}$-$N$) spin-independent 
cross sections as function of WIMP mass ($m_{\chi}$) at 90\% confidence level (C.L.) are derived using the dark matter halo model.
Within the context of the standard halo model,
the 90\% C.L. allowed regions implied by the DAMA/LIBRA and CoGeNT AM-based analysis are excluded at $>$99.99\% and 98\% C.L., respectively.
These results correspond to the best sensitivity at $m_{\chi}$$<$6$~{\rm GeV}/c^2$ among WIMP AM measurements to date.
\begin{description}
\item[PACS numbers]{95.35.+d, 29.40.-n, 98.70.Vc}
\end{description}
\end{abstract}

\maketitle


Compelling cosmological evidence indicates that about one-quarter of the energy density of the Universe 
manifests as dark matter~\cite{RPP-2018-review}, a favored candidate of which is the weakly interacting 
massive particle (WIMP, denoted as $\chi$). In direct laboratory searches of WIMPs conducted with 
WIMP-nucleus (${\chi}$-$N$) elastic scattering, positive evidence of WIMPs can only be established by 
assuming detailed knowledge of the background. The annual modulation (AM) analysis, on the other hand, 
only requires the background at the relevant energy range is stable with time. 
It can provide smoking-gun signatures for WIMPs independent of background modeling.
Within the astrophysical dark matter halo model~\cite{Drukier:1986}, the expected ${\chi}$-$N$ rates have distinctive AM features with maximum intensity in June and a period of 1 yr due to the Earth's motion relative to the galaxy dark matter distribution.

Positive results were concluded at significance of 12.9 $\sigma$ and 2.2 $\sigma$ from AM-based analysis 
of DAMA/LIBRA~\cite{Bernabei:2010,Bernabei:2013,Bernabei:2018yyw} and CoGeNT~\cite{Aalseth:2011b,Aalseth:2013,Aalseth:2014} experiments, 
respectively. However, these interpretations are challenged by integrated rate experiments with liquid 
xenon~\cite{Aprile:2018,Akerib:2017,Cui:2017}, cryogenic bolometer~\cite{Agnese:2014,Agnese:2016,Angloher:2016} 
and ionization germanium~\cite{Zhao:2013,Yue:2014a,Zhao:2016,Yang:2018b,Jiang:2018} detectors, 
when the data were analyzed in certain scenarios where the dark matter particle properties and distributions in the Milky Way's halo are precisely defined. Comparison of AM data with differnet targets is also model dependent. 
The AM-allowed regions of DAMA/LIBRA and CoGeNT have been probed and excluded by AM analysis from 
the XMASS-1 experiment~\cite{Abe:2018,Kobayashi:2019},
which is limited by the diminishing sensitivities of the liquid xenon techniques at light WIMP masses ($m_\chi$) below 6$~{\rm GeV}/c^2$.
The ANAIS-112~\cite{ANAIS-112:2019} and COSINE-100~\cite{COSINE-100:2019} experiments aim to resolve this tension by a model-independent test of DAMA/LIBRA's observation using identical detector target materials. Their latest results are consistent with both the null hypothesis and DAMA/LIBRA's 2$-$6 keV best-fit value, but at poor confidence levels due to the limited 1.5 yr and 1.7 yr data.
The CDEX experiment, located in the China Jinping Underground Laboratory (CJPL) with about 2400 m of rock overburden~\cite{cjpl:ARNPS17}, 
utilizes $p$-type point contact germanium detectors (PPCGe)~\cite{Luke:1989,Barbeau:2007,Soma:2016} for dark matter direct detection.
The low analysis threshold of about 200 eVee (``eVee'' represents electron equivalent energy derived from calibrations with known cosmogenic x-ray peaks)~\cite{Zhao:2013, Yue:2014a, Zhao:2016, Yang:2018b, Jiang:2018} implies AM studies with germanium can complement the liquid xenon results.
It provides an alternative probe to the allowed parameter space of DAMA/LIBRA~\cite{Bernabei:2010,Bernabei:2013} (with model dependence due to different target isotopes) and CoGeNT~\cite{Aalseth:2014} (with a model-independent comparison, since both use germanium as target) and extends  the reach of AM test to lower $m_\chi$.

The CDEX-1B experiment is the second phase of the CDEX experiment and has previously set upper limits for 
spin-independent (SI) and spin-dependent cross sections by the ${\chi}$-$N$ recoil spectral 
analysis~\cite{Yang:2018b}. The PPCGe target of mass 1 kg (fiducial mass of 939 g, after corrections due to 
a 0.88$\pm$0.12 mm surface layer) was shielded, from inside out, with 20 cm of copper, 20 cm of borated 
polyethylene and 20 cm of lead. The whole setup was assembled inside a 6 m ($H$)$\times$8 m ($L$)$\times$4 m ($W$) 
polyethylene room with wall thickness of 1 m. The target was enclosed by an NaI(Tl) anti-Compton detector 
from September 27, 2014 to August 2, 2017 (Run 1), and subsequently without NaI(Tl) (replaced by passive 
copper shielding) from August 4, 2017 till December 2, 2018 (Run 2). The gaps from December 27, 2014 to 
March 8, 2015 and from March 16, 2016 to June 2, 2016 were due to calibration with neutron and 
gamma-ray sources, respectively. The two runs have 751.3 and 428.1 live days, respectively, and together 
span a total of 1527 calendar days ($\sim$4.2 yr), with the total exposure of 1107.5 kg d.
The Run 1 events were further categorized by $\rm{AC}^{-(+)}$ corresponding to those without(with) coincidence of 
NaI(Tl) signals. Candidate ${\chi}$-$N$ events were therefore $\rm{AC}^{-}$ in Run 1 and all triggered ones in Run 2, 
which will also be denoted with $\rm{AC}^{-}$ in the following text for convenient purpose.
The energy calibration during the running period was achieved using the low energy internal x-rays 
from the cosmogenic nuclides inside the germanium crystal, also showing good stabilities.
The nuclear recoil spectral analysis of Run 1~\cite{Yang:2018b} achieved an analysis threshold of 
160~eVee, limited by the pedestal of the electronic noise. For AM analysis, good stability of contaminations 
due to electronic noise is required. Accordingly, a conservative analysis threshold of 250 eVee away from the 
pedestal noise edge is adopted, such that both the physics event selection efficiency and trigger efficiency 
are 100\%.

\begin{figure}[!hpbt]
\includegraphics[width=\linewidth]{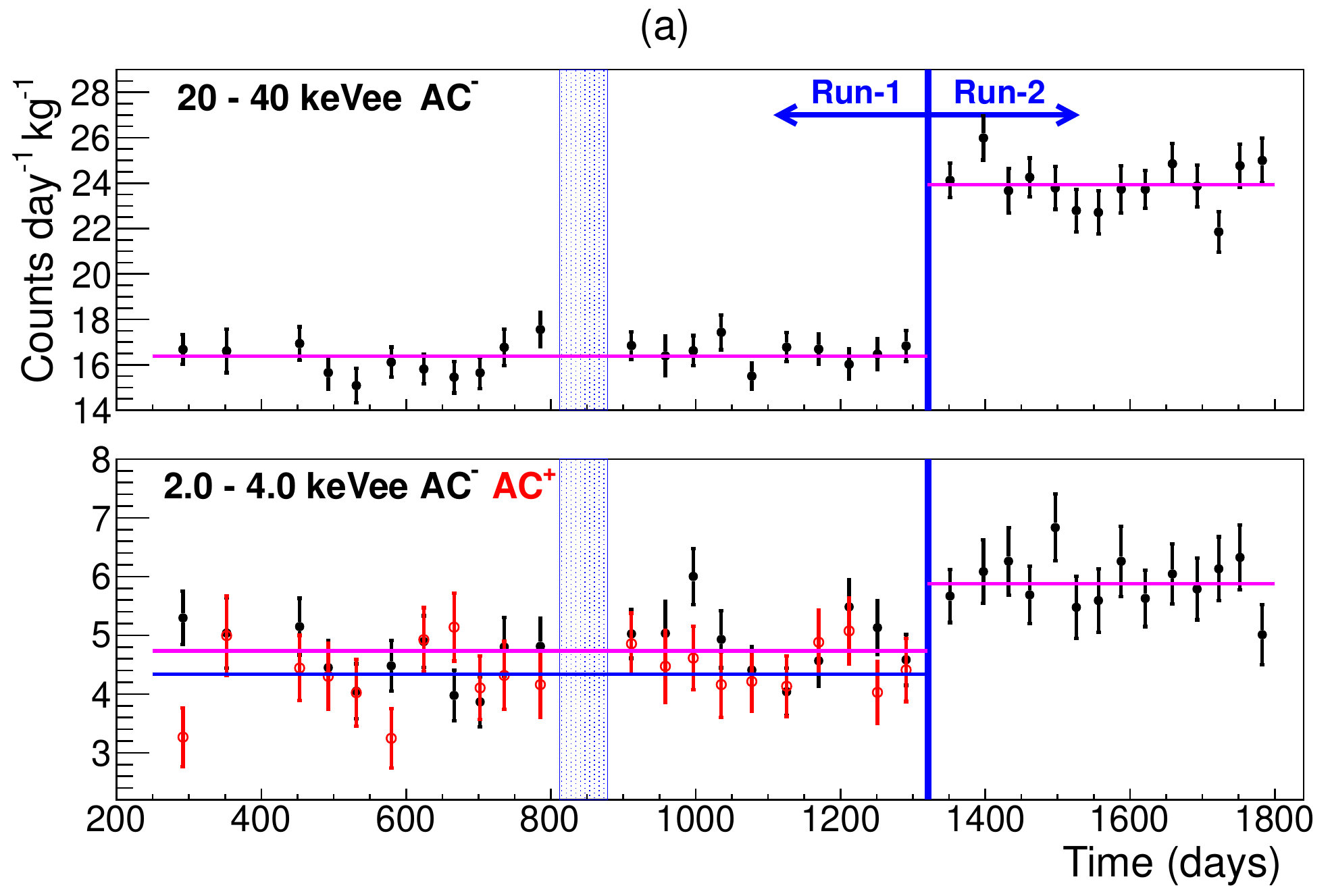}\\
\includegraphics[width=\linewidth]{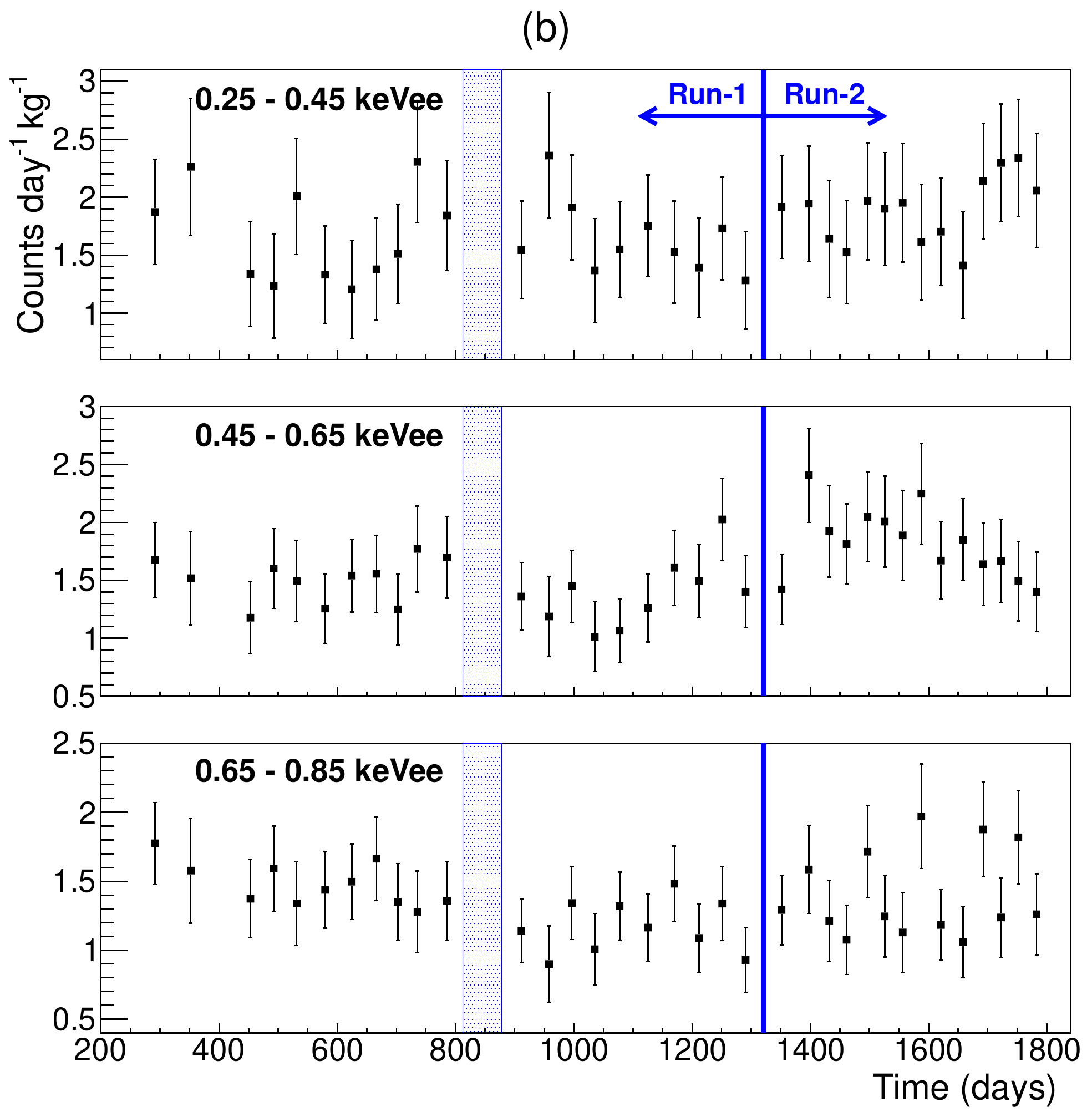}
\caption{
Count rates of CDEX-1B as function of time, where the shaded area denotes the period of gamma source calibration.
{\bf (a)} Time variation of the background count rates at 20$-$40 and 2.0$-$4.0 keVee.
Stability in the signal regions is demonstrated with good $\chi^2/\rm d.o.f.$ and $p$ values under 
stable-background hypothesis---17.30/20(0.63) for $\rm{AC}^{-}$ in Run 1, 16.44/13(0.23) for Run 2 at 20$-$40 keVee, 
and 28.15/20(0.11) for $\rm{AC}^{-}$ in Run 1, 18.66/20(0.54) for $\rm{AC}^{+}$ in Run 1, 9.24/13(0.75) for Run 2 at 2.0$-$4.0 keVee.
{\bf (b)} The B/S corrected bulk event counts versus time at three energy ranges which are most relevant to the sensitivities 
at $m_{\chi}$$\sim$8~${\rm GeV}/c^2$, with the overall $\chi^2/\rm d.o.f.$ of 35.5/60 for Run 1 and 24.6/39 for Run 2 in these energy ranges.
The uncertainties in these energies are dominated by the systematic uncertainties from B/S correction, 
in which overestimated upper bound was used~\cite{Yang:2018b}.
A bin size of 200 eVee is used in this plot, while different bin size is adopted in the analysis.
}
\label{fig:stability}
\end{figure}

At the keVee energy range relevant to this analysis, background events are dominated by Compton scattering 
of high energy gamma rays and by internal radioactivity from cosmogenic long-lived isotopes, 
the time variations of which have to be checked and accounted for. The time evolutions of radon contamination 
show good stabilities by the combined intensities of several 
radon-related $\gamma$ lines (295.2 and 351.9 keV from $^{214}$Pb, the daughter of $^{222}$Rn). 
The stabilities of the relevant background at the low energy are demonstrated in Fig.~\ref{fig:stability}(a), 
with the count rates at 20$-$40 and 2.0$-$4.0 keVee both for $\rm{AC}^{-}$ and $\rm{AC}^{+}$.
Time is denoted as the number of days since January 1, 2014.
It can be seen from the displayed $\chi^2/\rm d.o.f.$(degrees of freedom)
and $p$ values that the low energy background count rates are stable 
within the data taking periods.

The 4.2 yr of CDEX-1B data are separated into 35 subdatasets in different time bins, each with about 1 month of live time. 
WIMP candidate events in the bulk of the detector are selected~\cite{Yang:2018b} via some basic cuts and the bulk or surface (B/S) events discrimination.
The B/S correction procedure is done by likelihood fitting of the bulk or surface rise-time distribution probability density functions (PDFs) and has no cut efficiency associated, as described in details in Refs.~\cite{Yang:2018b,Yang:2018a}. 
During the B/S procedure, each subdataset was treated independently with its distinct calibration parameters.
The inputs of B/S procedure include (i) $\rm{AC}^{-}$ events in corresponding subdataset;
(ii) summation of $\rm{AC}^{-}$ in the rest of subdatasets; (iii) all $\rm{AC}^{+}$ in Run 1; and 
(iv) three calibration samples ($^{60}$Co, $^{137}$Cs, $^{241}$Am), while $^{241}$Am is a pure surface source and can supply the constrain to surface PDFs.
Systematic uncertainties related to the B/S correction are adopted from Ref.~\cite{Yang:2018b} and are combined quadratically.
The B/S corrections are stable within data taking periods, as checked by the stability of a few control parameters such as rise-time PDFs of background data, the counts of $\rm{AC}^{+}$ bulk and $\rm{AC}^{-}$ surface events.

The only requirement for AM analysis is to have stable background with time. The modeling of their origins 
and spectral shapes, which are sources of uncertainties in the time-integrated spectral analysis, is not involved. 
Stability of ${\chi}$-$N$ candidate events with time is further demonstrated in Fig.~\ref{fig:stability}(b) with the 
bulk event count rates after B/S correction at three energy ranges which are most relevant to 
the sensitivities at $m_{\chi}$$\sim$8~${\rm GeV}/c^2$. 
The data at low energies show slight time-dependent features. However, those features are not universal to all energy ranges.
Based on the physical understanding on the background components,
we adopted a scenario of the time-independent background contribution plus an exponentially time-dependent background contribution from the $L$-shell x rays from cosmogenic isotopes, which is not fitted, but derived from the corresponding $K$-shell lines intensities behavior. The expected time dependence due to the cosmogenic origin background contributions was observed. It dominated the background in energy ranges of 1.0$-$1.4 keVee, especially in Run 1.
The time-independent background levels of every energy bin were taken as free parameters and were uncorrelated between Run 1 and Run 2 due to the different shielding configurations.
The unmodulated ${\chi}$-$N$ rates were treated as a component of the constant background in AM analysis.

Data at 0.25$-$5.8 keVee were analyzed, below the region of internal $K$-shell x rays. 
The selected energy bin sizes are 50, 100, and 200 eVee for measured energy 
at $<$0.8, 0.8$-$1.6, and $>$1.8 keVee, respectively, 
according to the requirements of statistical accuracy in B/S correction.
The corrected counts of bulk events are denoted by $n_{ijk}$ corresponding to the respective bin with $i, j, k$ = (energy, time, run).
There are in total $i=1-40$~energy bins,
with 35 time bins divided into $k=1-2$ runs ($j=1-21$ time bins for $k=1$, $j=1-14$ time bins for $k=2$) in this analysis.
For each of the $i$th energy bin, a minimum $\chi^2$ analysis was performed simultaneously, with
\begin{equation}\label{eq_chi0}
\small
\chi_{i}^{2}=\sum_{k{\in}{\rm{Run}}}^{2}\sum_{j{\in}{\rm{Time}}}^{N}
\frac{[n_{ijk}-P_{ijk}-B_{ik}-A_{i}{\rm cos}(\frac{2{\pi}(t_{j}-\phi)}{T})]^{2}}{{\Delta_{ijk}}^{2}},
\end{equation}
where $\phi$ and $T$ are, respectively, the modulation phase and period.
The period is fixed at 365.25 d (one yr) for all scenarios, whereas the phase is either taken as free parameter or fixed at 152.5 d 
as expected from the standard halo model.
$P_{ijk}$ is the time-varying background contributions of the $L$-shell x rays from cosmogenic long-lived isotopes 
such as $^{68}$Ge, $^{68}$Ga  and $^{65}$Zn, the intensities of which are fixed by the measured $K$ shell x rays at 8.5$-$10.8 keV, 
$B_{ik}$ is the background level, in which we adopted a time-independent background scenario,
and ${\Delta_{ijk}}^{2}$ are the combined statistical and systematic errors dominated by the B/S correction~\cite{Yang:2018b}.
The modulation amplitude $A_{i}$ is fixed to 0 for the null hypothesis and left unconstrained (positive or negative) for the modulation hypothesis.
Summation is performed over all of the $j$th time bins each at median time $t_{j}$ and the $k$th run.

\begin{figure}[!hpbt]
\includegraphics[width=\linewidth]{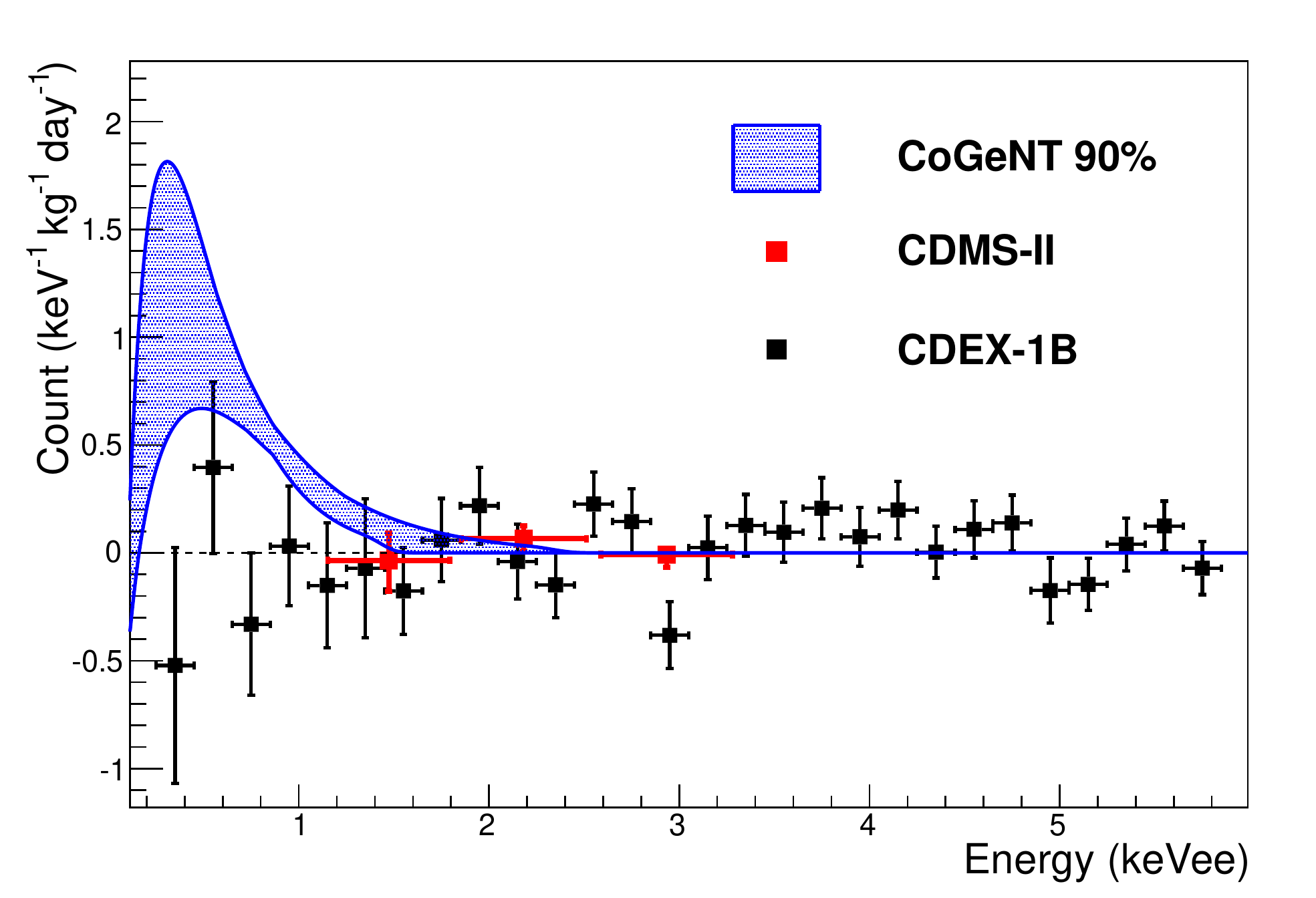}
\caption{
Best-fit solutions of modulation amplitude $A_{i}$ at phase=152.5 d.
The distributions show consistency with null results, i.e., no significant modulation signatures.
Derived modulation signals (based on standard halo model) from 90\% C.L. allowed regions of CoGeNT (3~yr)~\cite{Aalseth:2014}
and best-fit modulation amplitudes of CDMS-II distributed results~\cite{Ahmed:2012} are superimposed.
A bin size of 200 eVee was used for better illustration.
}
\label{fig:best_fit_Aik}
\end{figure}

The data are first studied with a model-independent analysis without invoking astrophysical models and parameters, i.e., model independently,
with the phase $\phi$ fixed at the halo-model expectation value of 152.5 d.
The modulation amplitudes $A_{i}$ of individual energy bins ($i$th) are treated independently,
from which the best-fit results with
$\chi^{2}/{\rm d.o.f.}=\sum\chi_{i}^{2}/{\rm d.o.f.}=1280.47/1280$ are shown in Fig.~\ref{fig:best_fit_Aik}.
The distribution of $A_{i}$ is consistent with null results, showing no evidence of modulation behavior.
These $A_{i}$ are contradicted with modulation amplitudes implied by the 90\% confidence level (C.L.) allowed region of CoGeNT~\cite{Aalseth:2014} at $p$ value$<$0.005.
The null hypothesis test gave a $\chi^{2}/\rm d.o.f.=1330.27/1320$.
The difference in $\chi^{2}$ between null hypothesis and independent-amplitude analysis is within 
$\chi^{2}$-distribution of $\rm d.o.f.$ of 40 (number of $A_{i}$) at $p$ value=0.14.

For the model-dependent analysis, the individual $A_{i}$ are correlated
with a known function ($f$) of $m_{\chi}$ and ${\chi}$-$N$ cross section, 
while the function is related to the applied astrophysics models.
The data are then analyzed under the standard spherical isothermal galactic halo model~\cite{Freese:2013,Drukier:1986}, 
with a most probable speed of $\upsilon_{0}=220$~km/s, a galactic escape velocity of $\upsilon_{\rm esc}=544$~km/s~\cite{Smith:2007}, 
an Earth's velocity related to dark matter of $\upsilon_{E}=\{232+15{\rm cos}~{2{\pi}(t-\phi)/T}\}$~km/s 
and local dark matter density of 0.3~${\rm GeV}/(c^2~{\rm cm^3)}$~\cite{Lewin:1996,Donato:1998}. 
The period and phase are fixed at 365.25 and 152.5 d, respectively.
Quenching factor of Ge is derived by the TRIM software package~\cite{Lin:2009,Ziegler:2004,Soma:2016} with a 10\% systematic error adopted for the analysis~\cite{Zhao:2016}.
Possible dark matter contributions which are not time varying are incorporated as part of $B_{ik}$.
The AM amplitudes $A_{i}$ are calculated by integration of $f$ with mean energy of the bin $E_{i}$ and bin size $\delta{E}_{i}$, that is,
$A_{i}=\sigma_{\rm SI}(m_{\chi}) f(E_{i}, \delta{E}_{i}; m_{\chi})$, where $\sigma_{\rm SI}$ denotes SI $\chi$-$N$ cross section as function of $m_{\chi}$.
Best-fit values of $\sigma_{\rm SI}$ are then evaluated by minimizing $\sum\chi_{i}^{2}$ of Eq.~(\ref{eq_chi0}).
The unified approach~\cite{Feldman:1998} is then used to place the upper bounds of positive definite $\sigma_{\rm SI}$ at different $m_{\chi}$.

\begin{figure}[!hpbt]
\includegraphics[width=\linewidth]{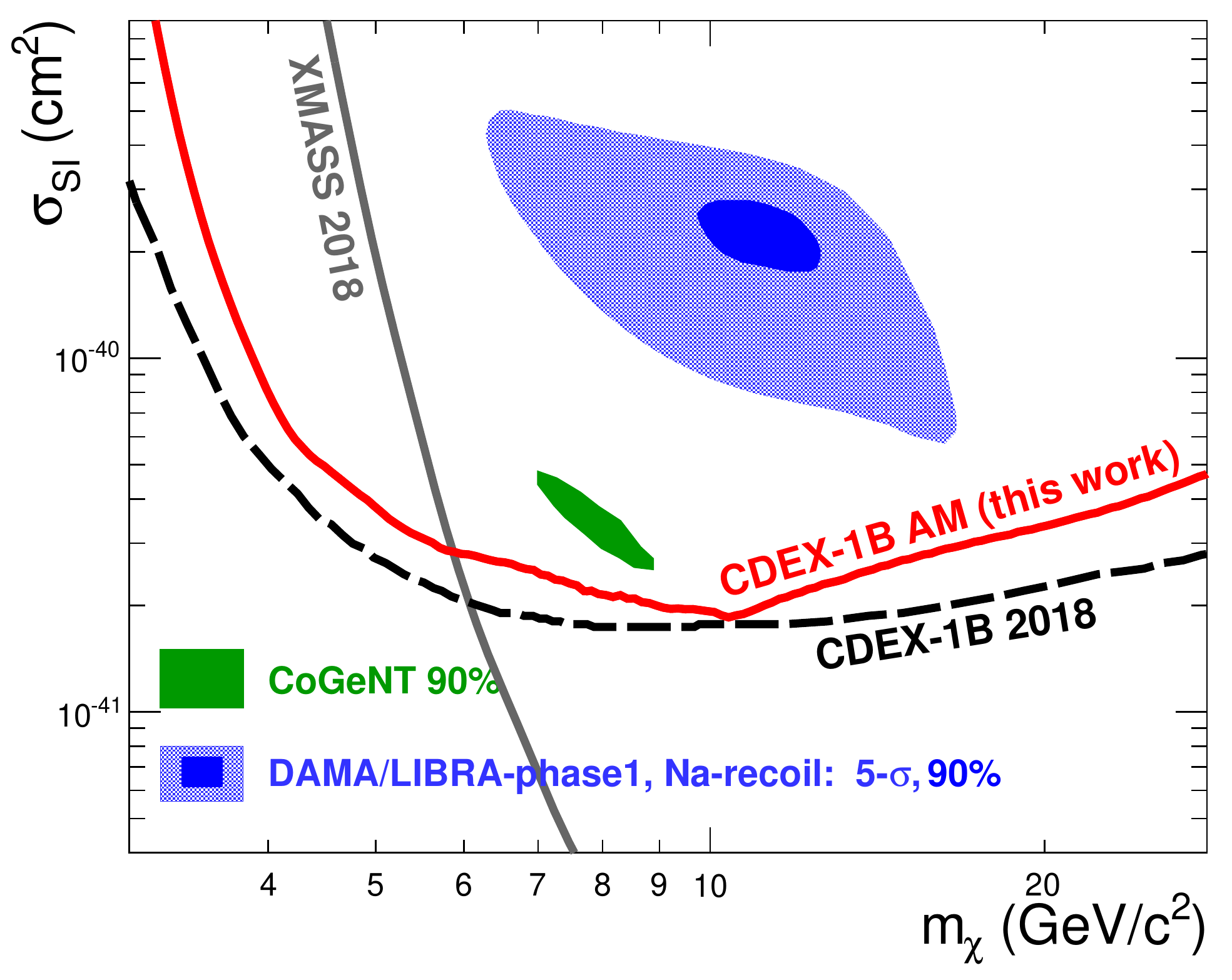}
\caption{
Limits at 90\% C.L. from CDEX-1B AM-analysis (red) on spin-independent WIMP-nucleon cross section. Also 
shown are other AM-based results: 90\% C.L. upper limits of XMASS-1 (dark gray)~\cite{Kobayashi:2019}, allowed regions of DAMA/LIBRA phase1
(Na-recoil, pale blue: 5-$\sigma$, blue: 90\% C.L.)~\cite{Savage:2009,Bernabei:2010,Bernabei:2013}, and 
CoGeNT (green: 90\% C.L.)~\cite{Aalseth:2014}. Constraints from the CDEX-1B time-integrated spectral 
analysis~\cite{Yang:2018b} are displayed (black dotted line) as comparison.
We note that the DAMA/LIBRA regions shown in this plot stem from the previous DAMA/LIBRA-phase1 data~\cite{Bernabei:2010,Bernabei:2013}, 
under the assumption of canonical isospin-conserving spin-independent WIMP-nucleus scattering in the standard halo model. 
New DAMA/LIBRA-phase2 data~\cite{Bernabei:2018yyw} have considerable impact on the best-fit regions and disfavor the canonical assumption~\cite{Baum:2018ekm}.
}
\label{fig:ex_plot}
\end{figure}

At $m_{\chi}$=7.9~${\rm GeV}/c^2$, the central value of $m_{\chi}$ of CoGeNT's 90\% C.L. allowed region ~\cite{Aalseth:2014}, the best-fit solution is
$\sigma_{\rm SI}=(-0.37{\pm}1.43){\times}10^{-41}~{\rm{cm}}^{2}$ ($\chi^{2}/\rm d.o.f.=1330.20/1319$), 
or equivalently, $\sigma_{\rm SI}<1.99{\times}10^{-41}~{\rm{cm}}^{2}$ at 90\% C.L.
The upper limits at 90\% C.L. on $\sigma_{\rm SI}$ are derived and shown in Fig.~\ref{fig:ex_plot}. The 
results refute the 90\% C.L. allowed regions inferred from AM-based analysis of 
DAMA/LIBRA-phase1 low-$m_{\chi}$ (Na-recoil)~\cite{Savage:2009,Bernabei:2010,Bernabei:2013} and 
CoGeNT~\cite{Aalseth:2014} experiments, providing an exclusion at $>$99.99\% and $>$98\% C.L., 
respectively. The DAMA/LIBRA high-$m_{\chi}$ region (I-recoil) is not probed in this analysis.

Systematic uncertainties on time-dependent background assumption are assessed by replacing constant backgrounds with linear functions, resulting in at most 3.4\% deviation of the upper bound and best fit of $\sigma_{\rm SI}$ for $m_{\chi}$ ranging from 2 to 20 GeV. The B/S discrimination contributes less than 8\% deviation of $\sigma_{\rm SI}$ and the uncertainty of $K/L$ ratios~\cite{Bahcall:1963} is also incorporated in the systematic uncertainty budget.

\begin{figure}[!hpbt]
\includegraphics[width=\linewidth]{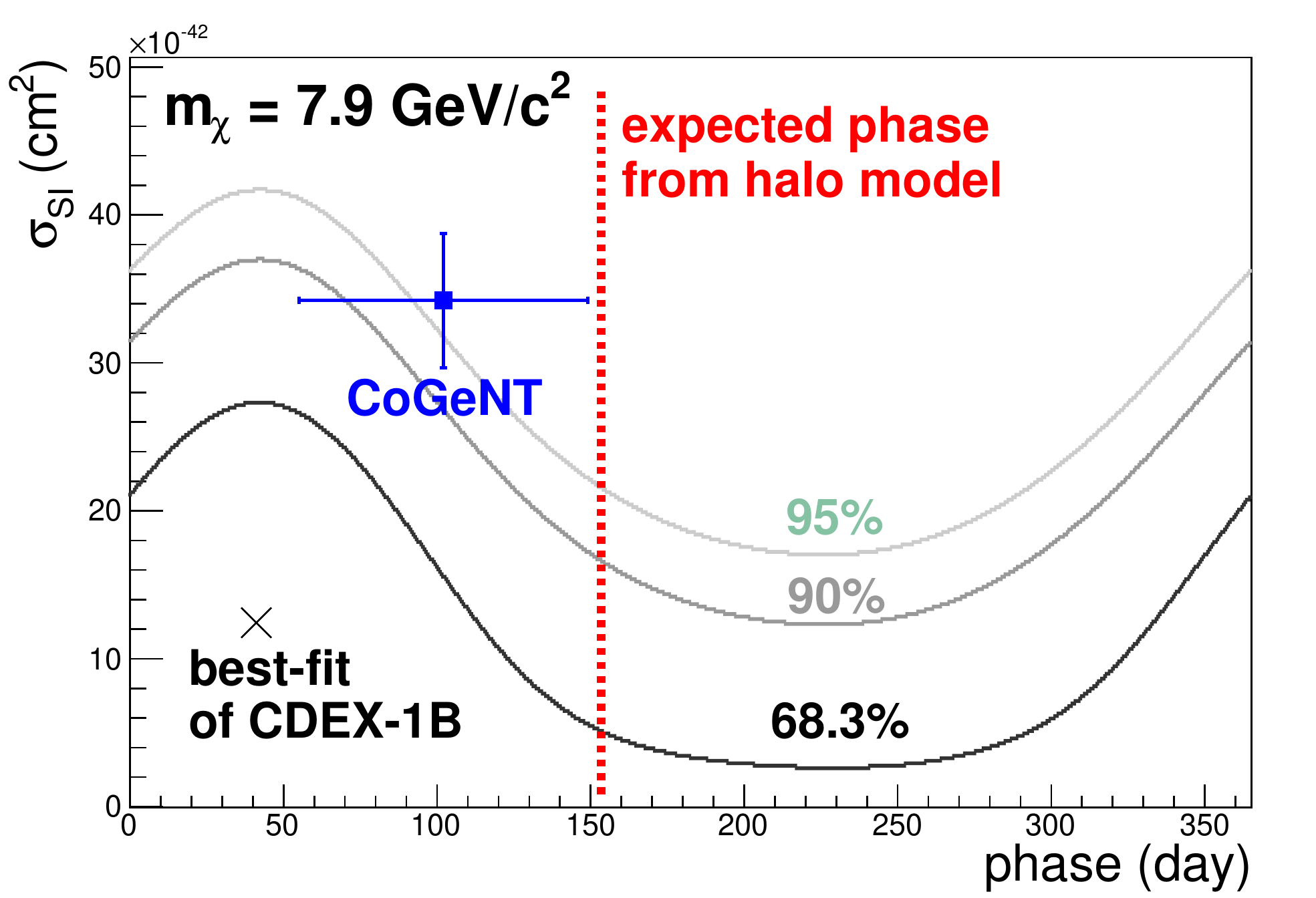}
\caption{
Best-fit $\sigma_{\rm SI}$ values versus $\phi$ at $m_{\chi}$=7.9~${\rm GeV}/c^2$, 
taking $\phi$ as a free parameter in $\sum\chi_{i}^{2}$.
The results show that the null hypothesis is within 1-$\sigma$ to the best-fit $\sigma_{\rm SI}$ values 
(similar conclusions also apply for $m_{\chi}$ in the range 3.2$-$17~${\rm GeV}/c^2$).
}
\label{fig:phase}
\end{figure}

The analysis is extended by taking the modulation phase $\phi$ as a free parameter,
and the exclusion contours of the best-fit results on $\sigma_{\rm SI}$ at $m_{\chi}$=7.9~${\rm GeV}/c^2$ are 
depicted in Fig.~\ref{fig:phase}, superimposed with the best-fit result from CoGeNT~\cite{Aalseth:2013, 
Aalseth:2014} at the same $m_{\chi}$ and the phase in halo model. 
The data exclude CoGeNT's 90\% C.L. allowed region at its best-fit phase of $102\pm47$ d~\cite{Aalseth:2014} and the halo model at fixed $\phi$ at 93\% and 98\% C.L., respectively.
The analysis at $m_{\chi}$ in the range 3.2$-$17~${\rm GeV}/c^2$ indicates that the data are consistent with the null hypothesis
within 1-$\sigma$ ($p$ value$>$0.32) at the entire $\phi$ range of $\phi$ from 0 to $2\pi$.

The CDEX-1B experiment provided unique low threshold (250 eVee) and stable (3.2 yr of live time) data 
for sensitive AM analysis results without energy-dependent background model assumptions. The CDEX dark matter program 
continues taking data at CJPL, expanding to use Ge-detector arrays immersed in liquid nitrogen acting 
as cryogenic coolant and shield against ambient radioactivity~\cite{Jiang:2018}. R\&D efforts on the 
Ge-detector fabrication, and further radiation background reduction, are being pursued. Scaled-up 
experiment toward target mass of 100 kg is being prepared at CJPL-II~\cite{cjpl:ARNPS17}.

This work was supported by the National Key Research and Development Program 
of China (Grant No. 2017YFA0402201) and the National Natural Science Foundation of 
China (Grants No. 11475092, No. 11475099, No. 11505101, No. 11675088, and No. 11725522), the Tsinghua University Initiative Scientific Research Program (Grant No. 20197050007), and the Academia Sinica Principal 
Investigator Award No. AS-IA-106-M02.

L. T. Y. and H. B. L. contributed equally to this work.
The authors of affiliations 2, 3 and 12 participated as members of TEXONO Collaboration.

\bibliography{cdex1b_am_v2.bib}

\end{document}